\documentclass[prl,twocolumn,superscriptaddress,floats]{revtex4}

\usepackage[dvips]{graphicx}
\usepackage{amsmath}
\usepackage{booktabs}
\usepackage{dcolumn}

\usepackage{rotating}

\newlength{\La} 

\newlength{\Lb} 

\newlength{\Lc} 

\newcolumntype{d}{D{.}{.}{-1}}

\newcommand{\bkbo}{Ba$_{1-x}$K$_x$BiO$_3$}

\newcommand{\lco}{La$_{2}$CuO$_4$}

\newcommand{\srruo}{Sr$_2$RuO$_4$}

\newcommand{\ybco}{YBa$_{2}$Cu$_{3}$O$_7$}
\newcommand{\hgbco}{HgBa$_{2}$CuO$_4$}
\newcommand{\ybcoa}{YBa$_{2}$Cu$_{3}$O$_{6+\delta}$}
\newcommand{\ybcos}{YBa$_{2}$Cu$_{3}$O$_{6}$}

\newcommand{\ncco}{Nd$_{2-x}$Ce$_{x}$CuO$_4$}
\newcommand{\nco}{Nd$_{2}$CuO$_4$}
\newcommand{\lsco}{La$_{2-x}$Sr$_{x}$CuO$_4$}
\newcommand{\lsmo}{La$_{1-x}$Sr$_{x}$MnO$_3$}
\newcommand{\lanio}{La$_{2}$NiO$_4$}

\newcommand{\nccofz}{Nd$_{1.85}$Ce$_{0.15}$CuO$_4$}
\newcommand{\nccoef}{Nd$_{1.85}$Ce$_{0.15}$CuO$_4$}
\newcommand{\lscofz}{La$_{1.85}$Sr$_{0.15}$CuO$_4$}
\newcommand{\lscoef}{La$_{1.85}$Sr$_{0.15}$CuO$_4$}

\newcommand {\vQ}{$\mathbf{Q}$}

\newcommand {\Eu}{$E_{u}$}
\newcommand {\de}{$\Delta_{1}$}
\newcommand {\se}{$\Sigma_{1}$}
\newcommand {\De}{$\Delta_{1}$}

\newcommand {\vZ}{$\mathbf{Z}$}

\newcommand {\vq}{$\mathbf{q}$}

\newlength{\figwidth}

\divide\figwidth by 2

\begin{document}

\advance\vsize by 2 cm

\title{Dispersion of the high-energy phonon modes in Nd$_{1.85}$Ce$_{0.15}$CuO$_4$.}

\author{M.~ Braden}
\email{braden@ph2.uni-koeln.de}%
\affiliation{
II. Physikalisches Institut, Universit\"at zu K\"oln,
Z\"ulpicher Str. 77, D-50937 K\"oln, Germany}

\author{L.~ Pintschovius}
\affiliation{Forschungszentrum Karlsruhe, Institut für
Festkörperphysik, P.O.B. 3640, D-76021 Karlsruhe, Germany}
\affiliation{Laboratoire L\'eon Brillouin, CE Saclay, F-91191
Gif-sur-Yvette, France}

\author{T.~ Uefuji}
\affiliation{ Institute for Chemical Research, Kyoto University,
Uji 611-0011, Japan}

\author{K.~ Yamada}
\affiliation{Institute of Materials Research, Tohoku University,
Sendai 980-8577, Japan}

\date{\today, \textbf{preprint}}

\pacs{PACS numbers: }

\begin{abstract}

The dispersion of the high-energy phonon modes in the electron
doped high-temperature superconductor
Nd$_{1.85}$Ce$_{0.15}$CuO$_4$ has been studied by inelastic
neutron scattering. The frequencies of phonon modes with Cu-O
bond-stretching character drop abruptly when going from the
Brillouin zone center along the [100]-direction; this dispersion
is qualitatively similar to observations in the hole-doped
cuprates. We also find a softening of the bond-stretching modes
along the [110]-direction but which is weaker and exhibits a
sinusoidal dispersion. The phonon anomalies are discussed in
comparison to hole-doped cuprate superconductors and other
metallic perovskites.

\end{abstract}

\maketitle

\section{I. Introduction}

The question about the role of electron phonon coupling in
high-temperature superconductivity is still matter of controversy
\cite{andersen,shen,lanzarra,falter,tachiki-neu} and, most likely,
will not be solved in the nearest future. Therefore, it appears
interesting to continue to look for direct evidence for electron
phonon coupling in these materials. In the conventional
superconductors \cite{3}  signatures of strong electron phonon
coupling can be found either in the form of dips in the phonon
dispersion or by comparison with a non-superconducting reference
material. In the case of the cuprates, however, there are no
metallic non-superconducting reference materials; therefore, a
safer way to discern electron-phonon  coupling effects is to
compare the phonon dispersion curves of cuprate superconductors
to those of their insulating parent compounds. The latter
approach is adopted in this study.

Very early, evidence for strong electron phonon coupling in the
cuprate high-temperature superconductors was found in inelastic
neutron scattering (INS) experiments on \lsco ~ and \ybcoa \
\cite{4}. The comparison of the phonon density of states (PDOS)
measured on the insulating parent compounds, \lco ~ and \ybcos ,
with that obtained on the optimally doped superconducting
materials, \lscofz ~ and \ybco , reveals a pronounced shift of
spectral weight. Doping induces a frequency renormalisation of
the modes with the highest energies which possess a longitudinal
bond-stretching character, see Fig. 1. In contrast, the low and
medium energy ranges in the PDOS exhibit less sensitivity to
doping. More recently, similar effects were also reported for the
electron doped superconducting cuprates \cite{ndcePDOS,kang}.

Studies of the phonon dispersion using single crystals and triple
axis neutron spectrometry have clarified the shift of weight in
the PDOS \cite{5}. Along the [100] direction, the branches of the
longitudinal bond-stretching character exhibit a softening when
going from the zone-centre into the zone, for the polarization
patterns of the bond-stretching modes see Fig. 1. Qualitatively,
the dispersion is similar in \lscofz ~\cite{6,7} and in \ybco
\cite{8,9,10}. The branches are nearly flat  close to the
Brillouin zone centre, then show a steep decrease in frequency
around \vq =(0.25 0 0) and become flat again between (0.35 0 0)
and (0.5 0 0).
Along the [110]
direction a nearly flat dispersion is observed in \lsco ~ and in
\ybco , which still represents sizeable frequency renormalisation
compared to the insulating parent compounds.
The strong frequency renormalization of the zone-boundary modes
is, however, not a particularity of the superconducting cuprates,
but is seen in many  doped perovskites : nickelates \cite{15,21},
manganates \cite{14} and superconducting \bkbo \cite{12,19,20},
the only metallic perovskite compound which does not exhibit the
effect is superconducting \srruo \cite{braden-unp}.


\begin{figure}[tp]
\resizebox{0.7\figwidth}{!}{
\includegraphics*{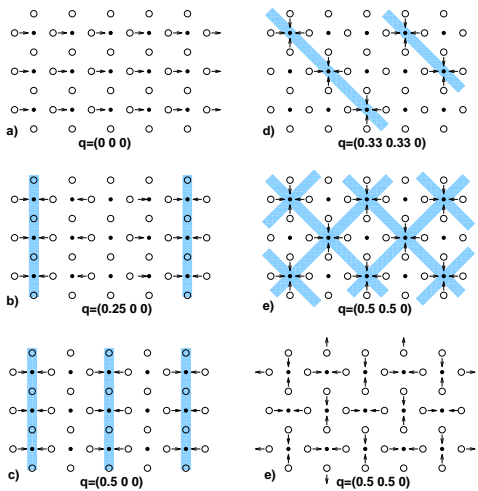}
} \caption{ Polarization patterns of the Cu-O bond-stretching
phonon modes for different propagation vectors : the zone center
mode (a), the longitudinal mode for \vq =(0.25,0,0) (b), the
linear or half-breathing mode (c), the longitudinal mode for \vq
=(0.33,0.33,0) (d), the planar breathing mode (e), and the
quadrupolar mode of transverse character at \vq =(0.5,0.5,0).
}\label{fig1}
\end{figure}

We note that that the softening of the longitudinal
bond-stretching phonons cannot be understood from
phenomenological lattice dynamical models like the shell model.
The insulator-to-metal transition upon doping entails a certain
frequency reduction as a result of screening effects suppressing
the LO-TO splitting of the polar modes at the Brillouin zone
centre. However, in materials like the cuprates screening effects
are confined to the long-wavelength phonons as it is illustrated
in Fig. 2. Such behavior is indeed observed in the dispersion of
the low-energy phonon branches with polar character in many oxide
perovskites \cite{12,13,14,15}.

\begin{figure}[tp]
\resizebox{0.4\figwidth}{!}{
\includegraphics*{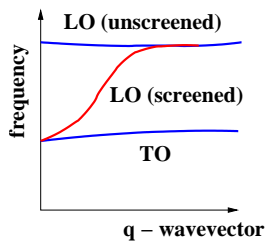}
} \caption{ Illustration of the screening of polar modes in a
metal with low carrier concentration, when the electronic
susceptibility is described in a free electron model.}\label{fig2}
\end{figure}

Until recently the analysis of the bond-stretching phonon
dispersion in the superconducting cuprates was restricted to \lsco
~ and \ybcoa , since only for these compounds it has been possible
to grow single crystals of sufficient size for inelastic neutron
scattering However, with the recent advance in synchrotron
sources, inelastic x-ray scattering became more competitive and
was applied to \ncco ~ \cite{26,27}, \hgbco \cite{28} and very
recently to \lsco \ \cite{fukuda}. For \hgbco ~ Uchiyama et al.
\cite{28} find a dispersion similar to that observed in \ybco \
\cite{8,9}, i.e. the same step-like downwards dispersion in the
bond-stretching branch along [100]. In the electron-doped
material \ncco ~ the high-energy dispersion is more complex since
there is another longitudinal branch of the same symmetry which
interacts with the bond-stretching branch, see below. d'Astuto et
al. \cite{26,27} report that the frequencies of the modes with
Cu-O bond-stretching character drop near \vq $\sim$(0.2~0~0)
falling below those of an other branch. However, the intensity
related to bond-stretching character could not safely be followed
beyond (0.15 0 0) in the inelastic x-ray scattering experiment.
This motivated us to investigate the high-energy phonon
dispersion in \nccofz ~ by inelastic neutron scattering after
large superconducting single crystals of this material became
finally available.

The paper is organized as follows: the next Section describes our
experimental technique. Thereafter, we explain the
phenomenological model used to analyze the data.  Then, we
present results for the [100] direction and the [110] direction
and discuss the polarization patterns of the different phonon
branches. Finally, we compare the phonon anomalies found in \ncco
~ to those observed in hole-doped cuprates and also to those
observed in other metallic oxide perovskite materials.

\section{II. Experimental}

Two single crystals of \nccofz ~ were grown by a floating zone
technique using a mirror furnace at Kyoto University. The crystals
exhibit the superconducting transition at 25\ K and samples
obtained with this method were already used in many studies
\cite{yamada-1,yamada-2}. Inelastic neutron scattering experiments
were performed on the 1T spectrometer at the Orph\'ee reactor
using Cu(111) and Cu(220) monochromators and a pyrolithic graphite
(002) analyzer. Higher order contaminations were suppressed with a
pyrolithic graphite filter in front of the analyzer. Double
focusing was applied on the monochromator and on the analyzer
sides. Two single crystals, each of about 500mm$^3$ volume, were
co-aligned in the [100]/[010] scattering geometry and cooled in a
closed He-cycle cryostate. An additional experiment was performed
with one crystal in the [100]/[001] orientation.

\begin{figure}[tp]
\resizebox{0.35\figwidth}{!}{
\includegraphics*{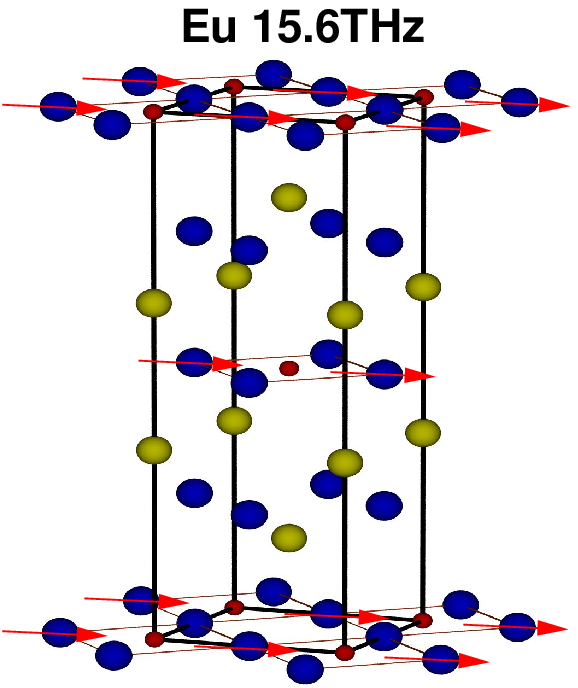}
} \resizebox{0.33\figwidth}{!}{
\includegraphics*{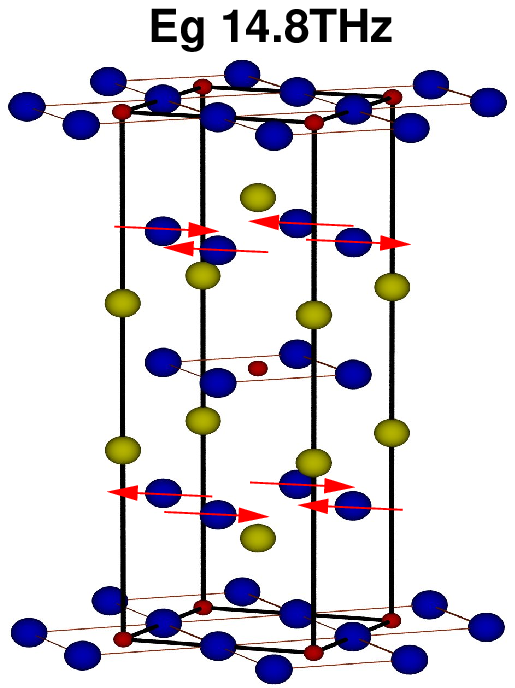}
} \resizebox{0.35\figwidth}{!}{
\includegraphics*{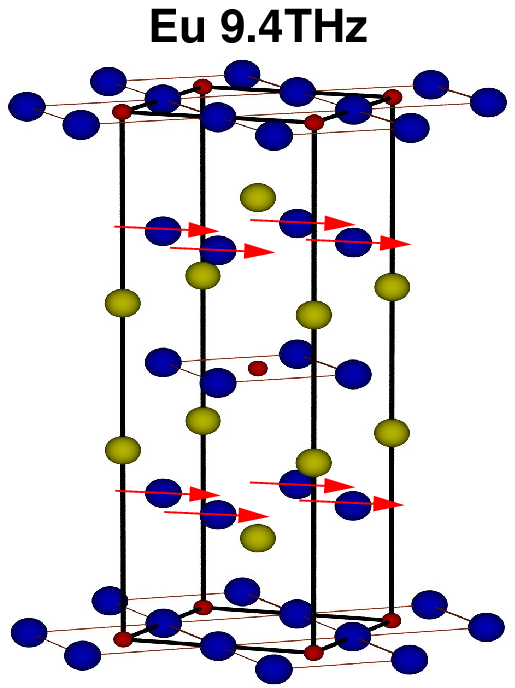}
} \resizebox{0.35\figwidth}{!}{
\includegraphics*{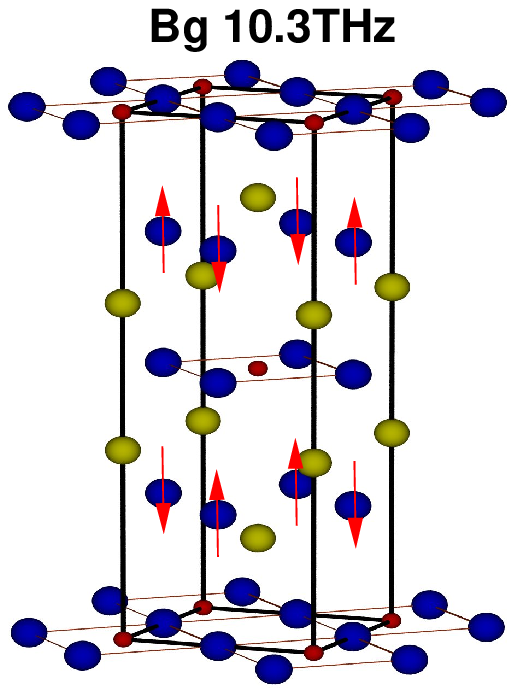}
} \caption{ Polarization patterns of the zone-center modes
relevant for the \de ~high-energy dispersion. Large blue balls
correspond to oxygen atom, small red balls to Cu atoms and the
large yellow ones to Nd/Ce-sites. }\label{fig3}
\end{figure}

\section{III. Results and discussion}

The polarization patterns of the  Cu-O plane-polarized
bond-stretching modes are depicted in Fig. 1. For the longitudinal
modes at the zone-boundaries there is an alternation of Cu-sites
with short and long metal-oxygen distances and therefore, the
modes at (0.5,0,0) and (0,5,0,5,0) are called half-breathing and
breathing modes, respectively. These modes may strongly couple to
the charges, since smaller metal-oxygen distances point to a
higher metal valence \cite{11}. Fig. 1 depicts also the
polarization patterns of some other plane-polarized Cu-O1
bond-stretching modes whith a regard to discussion which modes can
be expected to couple strongly to charge fluctuations. To this
end, loci of dynamical charge accumulation favored by the atomic
displacements are high-lighted.

\subsection{III.1  Phenomenological model }

The above mentioned anti-crossing phenomena - which are absent in
the case of \lsco ~ because of a different structure - makes the
use of a lattice dynamics model indispensable to correctly assign
the various high-energy phonon modes in \ncco . This interaction
of the longitudinal optical branches has been seen in the x-ray
studies \cite{26,27} and is predicted by calculations with the
phenomenological model. The complication arises from the shift of
the second oxygen site (O1 labels the in-plane oxygen and O2 the
out-of-plane one) from the apical position at (0 0 0.17) in \lsco
~ to (0.25\ 0.25\ 0.25) in \ncco \ \cite{30}, where it is stronger
bonded with the RE-ions. In consequence, the vibrations of O2
parallel to the CuO$_2$-planes exhibit higher energies and may
interact with the Cu-O bond-stretching modes in \ncco .
In order to put the assignment of the various high-energy modes
onto firm ground, we measured the dispersion of all high-energy
branches of a particular symmetry and compared the measured
scattering intensities with the model predictions. Since it was
beyond the scope of our experiments to determine  a full set of
dispersion curves we did not develop a new model but adapted the
one fitted to the phonon dispersion of the parent compound \nco
~which is described in reference \cite{29}. Only a few changes
were necessary in order to achieve a satisfying agreement with the
experimentally obtained dispersion. The model parameters are
given in Table II, the crystal structure data were taken from
reference \cite{30}. We use a shell model \cite{35} with
Born-Mayer potentials describing the repulsive interactions. The
shell charges of Nd and Cu and the amplitude of the O-O
Born-Mayer potential were slightly adapted. Further, we included
a longitudinal force constant between Cu and the O2-oxygen as was
done for the non-doped compound \cite{29} to arrive at a fully
satisfactory description of the experimental dispersion curves.

The main lattice dynamical difference between the parent and the
doped compounds concerns the screening of the Coulomb-interaction
through the free charge carriers.  We describe the electronic
susceptibility by the Lindhard-function for a free electron gas
taking exchange corrections into account. Such model invariably
produces complete screening at long distances, which means that
the splitting between longitudinal optic (LO) and transverse
optic (TO) phonon frequencies will disappear at the zone center.
However, when passing into the zone, the screening has to occur
at smaller distances where it will be less effective and the
LO-TO splitting may partially recover yielding an increasing
dispersion. This scenario describes perfectly the dispersion of
phonon branches with polar character at intermediate energies in
many doped perovskites \cite{12,13,14,15}. In particular, it
applies to the modes with the strongest polar character which are
the bond-bending modes related to ferroelectricity.

In the case of \ncco ~ we obtain a screening vector of $k_s$ =
0.42\AA$^{-1}$ in good agreement with the analysis of the
inelastic x-ray scattering results \cite{26}. The
screening-vector is furthermore comparable to the one found in
\lsco , $k_s$ = 0.39\AA$^{-1}$ for x$\sim$0.1 \cite{29}. In view
of the higher doping in our sample the screening in the electron
doped system appears to be somewhat less effective than in the
hole-doped compounds. However, in our analysis the screening is
mainly determined by the slope of the O2-branch. Therefore, we
cannot exclude that the screening is somewhat  more effective in
the Cu-O plane.

\bigskip
\begin{table}
\begin{tabular}{c c c c }
\multicolumn{4}{c}{ionic part} \\
\hline
ion   & Z     & Y     & K    \\
Cu    & 1.64  & 1.4   & 2.0  \\
Nd    & 2.30  & -4.9  & 8.0  \\
O1    & -1.56 & -3.0  & 2.0  \\
O2    & -1.56 & -3.0  & 2.0  \\
\end{tabular}
\begin{tabular}{c c c c}
\multicolumn{4}{c}{potentials} \\
\hline
pair  & A(eV)  & r$_0$ (\AA ) & C (eV/\AA $^6$ \\
Nd-O  &  2000  & 0.319 & -                  \\
Cu-O  &  3950  & 0.228 & -               \\
O-O   &  2650  & 0.284 &  -100              \\
~ & ~ & ~ & ~ \\
\end{tabular}
\begin{tabular}{c c }
\multicolumn{2}{c}{force constants (dyn/cm)} \\
\hline
pair  &  F       \\
Cu-O2 (long.) & 13372    \\
lin.-breathing & 93800      \\
plan.-breathing & 69300 \\
\end{tabular}
\nobreak \caption{Lattice dynamics model  parameters for the
description of the phonon dispersion in \nccofz ; for the
explanation of the parameters see text; Z,Y are given in electron
charges; K in $10^6$dyn/cm. }\label{tab2}
\end{table}

As has been pointed out in the introduction, screening treated
within the free electron model is unable to explain the downward
dispersion of the longitudinal branches along the [100] and the
[110] directions. In order to reproduce such a behavior - which
is sometimes called over-screening - we included additional force
constants lowering the frequencies of the Cu-O1 bond-stretching
modes at the zone boundary in the [100] (linear breathing term)
or in the [110] direction (planar breathing term), respectively.
The linear and planar breathing constants ( see Table I)
correspond to force constants between a particular O1-site and
the other O1's connected with the same Cu-atom. A displacement of
the first O1 towards the Cu-site results in forces on the other
O1's also directed towards the Cu-site. With the linear and
planar breathing constant, the forces act on just the opposite
and on all three surrounding O1's, respectively. With these
parameters a quantitative description of the dispersion is
achieved \cite{kommentar1}.
We note that such terms have been used
previously to describe the phonon anomalies in hole-doped
compounds (see, e.g., \cite{10}).

The softening of the zone-boundary modes observed in the
superconducting cuprates can be considered as being due to an
over-screening of the Coulomb-potentials requiring a special
electronic susceptibility \cite{16,17}. Assuming that charges may
fluctuate between different cation-sites on the frequency scale
of the phonon modes one may explain the softening of the
breathing modes \cite{falter,16,17}. Ab initio calculations of
the phonon dispersion for the hypothetic material CaCuO$_2$
\cite{savrasov} and for \ybco ~ \cite{18} agree with the soft
half-breathing mode frequency rather well, but they also obtain a
soft breathing mode at \vq =(0.5 0.5 0) and they do not find the
step-like dispersion \cite{savrasov,18}. Although the breathing
force constants in our model do not posses a clear physical
meaning they appear to mimic the particular electronic
susceptibility in the metallic cuprates \cite{kommentar1}.

\begin{table}
\begin{tabular}{| c | c |c |c |c |c  |c  |c  |}
\hline
~~ & Nd & Nd' & Cu  & O1 & O1' & O2 & O2'  \\
~~ & 0\ 0\ .35 & 0\ 0\ -.35 & 0\ 0\ 0 & .5\ 0\ 0 & 0\ .5\ 0&.5\ 0\ .25 & 0\ .5\ .25 \\
\hline
1\ B$_{g}$ & 0 0 0 & 0 0 0 & 0 0 0 & 0 0 0 & 0 0 0 & 0 0 A& 0 0 -A \\
1\ B$_{u}$ & 0 0 0 & 0 0 0 & 0 0 0 & 0 0 A & 0 0 -A & 0 0 0& 0 0 0 \\
1\ A$_{g}$ & 0 0 A & 0 0 -A & 0 0 0 & 0 0 0 & 0 0 0 & 0 0 0& 0 0 0\\
4\ A$_{u}$ & 0 0 A & 0 0 A & 0 0 B & 0 0 C & 0 0 C & 0 0 D& 0 0 D\\
2\ E$_{g}$ & A 0 0 & -A 0 0 & 0 0 0 & 0 0 0 & 0 0 0 & B 0 0& -B 0 0\\
5\ E$_{u}$ & A 0 0 & A 0 0 & B 0 0 & C 0 0 & C 0 0 & D 0 0& D 0 0\\
\hline
7 $\Delta _1$  & A 0 B & A 0 -B & C 0 0 & D 0 0 & E 0 0 & F 0 G& F 0 -G  \\
2 $\Delta _2$   & 0 A 0 & 0 -A 0 & 0 0 0 & 0 0 0 & 0 0 0 & 0 B 0& 0 -B 0 \\
5 $\Delta _3$   & 0 A 0 & 0 A 0 & 0 B 0 & 0 C 0 & 0 D 0 & 0 E 0& 0 E 0 \\
7 $\Delta _4$   & A 0 B & -A 0 B & 0 0 C & 0 0 D & 0 0 E & F 0 G& -F 0 G \\
\hline
6 $\Sigma _1$  & A A B  & A A -B & C C  0 & D E 0 & E D 0 & F F 0& F F 0 \\
3 $\Sigma _2$  & A -A 0 & -A A 0 & 0 0  0 & 0 0 B & 0 0 -B & C C 0& -C -C 0  \\
6 $\Sigma _3$  & A -A 0 & A -A 0 & B -B 0 & C D 0 & -D -C 0 & E -E F& E -E -F \\
6 $\Sigma _4$  & A A B  & -A -A B & 0 0 C & 0 0 D & 0 0 D & E -E F & -E E F \\
\hline

\end{tabular}

\begin{tabular}{| r @{~:~} l @{~~~~} r @{~:~} l |}
\hline
7\ $\Delta _1 $ &  5\ E$_u$~+~1\ A$_{g}$~+~1\ B$_{g}$ & 5\ $\Delta _3 $ &  5\ E$_{u}$ \\
2\ $\Delta _2 $ &  2\ E$_{g}$ & 7\ $\Delta _4$&  3\ E$_{g}$~+~4\ A$_{u} $~+~ B$_u$ \\
\hline
6\ $\Sigma _1 $ &  5\ E$_{u}$~+~1\ A$_{g}$ & 6\ $\Sigma _3 $ &  5\ E$_{u}$~+~B$_{g}$ \\
3\ $\Sigma _2 $ &  2\ E$_{g}$~+~ B$_{u}$ & 6\ $\Sigma _4 $ &  4\ A$_{u}$~+~2\ E$_{g}$ \\
\hline
\end{tabular}

\medskip
\caption{Upper part :Polarization schemes according to the crystal
structure of \ncco ~ for all $\Gamma$-modes and for the
representations along the [100] and [110] directions (labeled
$\Delta$ and $\Sigma$ respectively); the first line gives the
positions of 7 atoms forming a primitive unit, the following
lines show the displacements of these atoms. A letter at the
$i$-position, signifies that this atom is moving along the
$i$-direction, a second appearance of the same letter signifies
that the second atom moves with the same amplitude ("-" denotes a
phase shift) in the corresponding direction. Lower part :
compatibility relations along the [100] and [110] directions}
\end{table}

Table II resumes the group theoretical analysis. The zone center
modes can be separated according to the irreducible
representations, whose the polarization patterns are given in
Table II. Along the [100] ($\Delta$) and along the [110] direction
($\Sigma$), modes split into four different representations, all
compatibility relations are given in Table II.

\begin{figure}[tp]
\resizebox{0.9\figwidth}{!}{
\includegraphics*{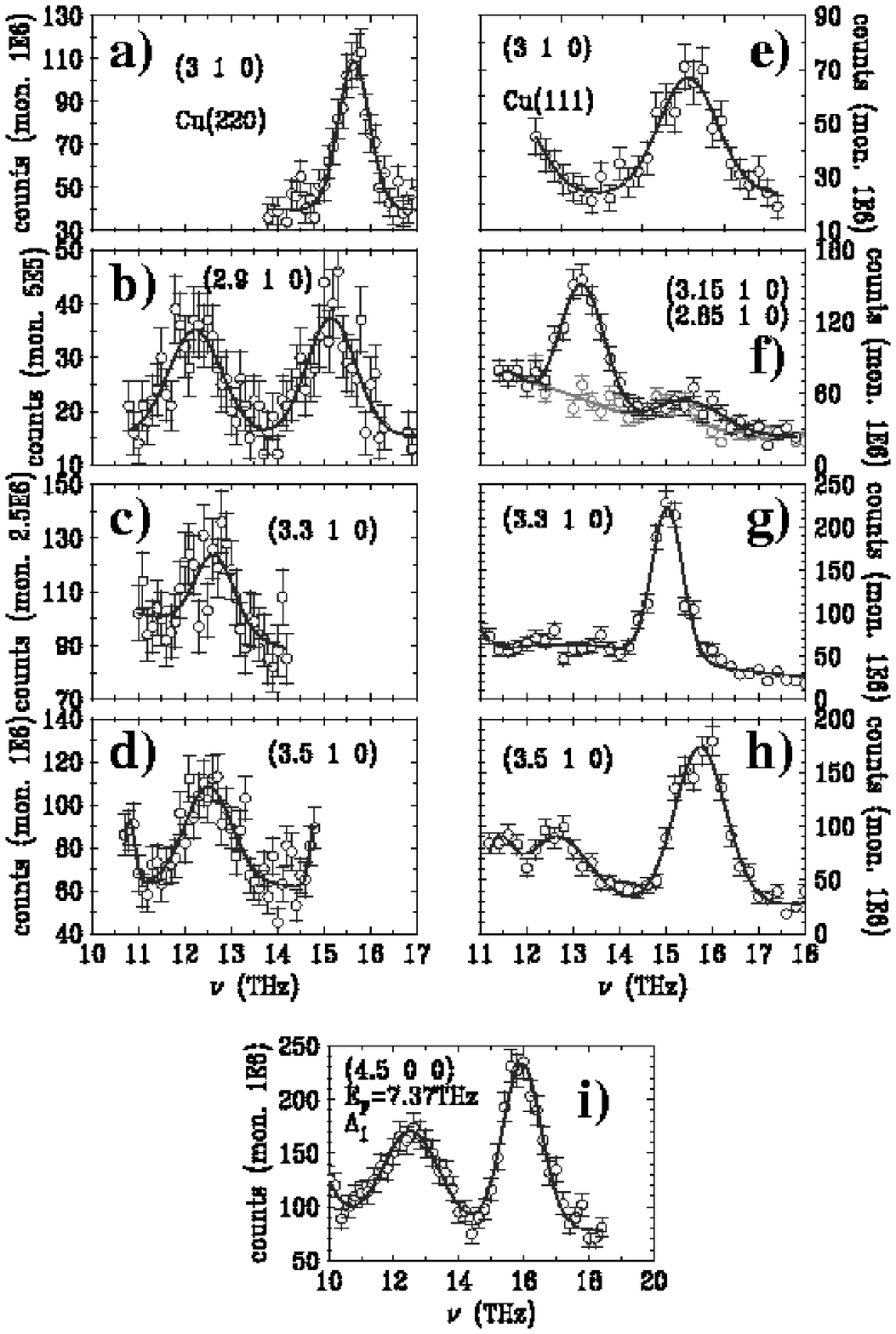}
}
\caption{ Scans across the longitudinal bond-stretching modes
with \de -symmetry: a)-d) are obtained with high resolution
conditions (Cu(220) monochromator), e)-h) with medium resolution
(Cu(111)-monochromator) and i) with low resolution
(Cu(111)-monochromator with a final energy of 7.37\ THz) . Lines
correspond to fits with Gaussians. }\label{scanx00}
\end{figure}

\subsection{III.2 Phonon dispersion in [100] direction }

In the [100] direction the branches corresponding to the \de
-representation start from the $E_u$-, $A_g$- and $B_g$-modes at
the zone center. The polarization patterns of the four modes with
the highest energies are shown in Fig. 3. The patterns were
calculated with the lattice dynamical model fitted to the measured
dispersion, see above. In addition to the CuO1 bond-stretching
mode (\Eu ) at the highest energy, there is the CuO1 bond-bending
mode (\Eu ) at about 10THz, the $B_g$-mode at 10.3\ THz and the
O2-mode (\Eu ) at 9.4\ THz. The $B_g$-mode corresponds to a motion
of O2 parallel to the $c$-axis; its frequency is observed by
Raman-scattering at 10.0--10.3\ THz \cite{31,32,33}. \Eu -modes
are Infra-Red active; however, the reported measurements are not
conclusive, possibly due to insufficient sample quality
\cite{33,34}. Fig. 3 shows also the displacement pattern of the
highest $E_g$ mode. This mode is the starting point of a branch
with $\Delta_4$ symmetry which lies in the same frequency range as
those associated with the Cu-O1 respectively Nd-O2 bond-stretching
vibrations. Measurements aiming at the dispersion of the latter
branches were performed in Brillouin zones with zero structure
factor for the $\Delta_4$-modes.

\begin{figure}[tp]
\resizebox{0.95\figwidth}{!}{
\includegraphics*{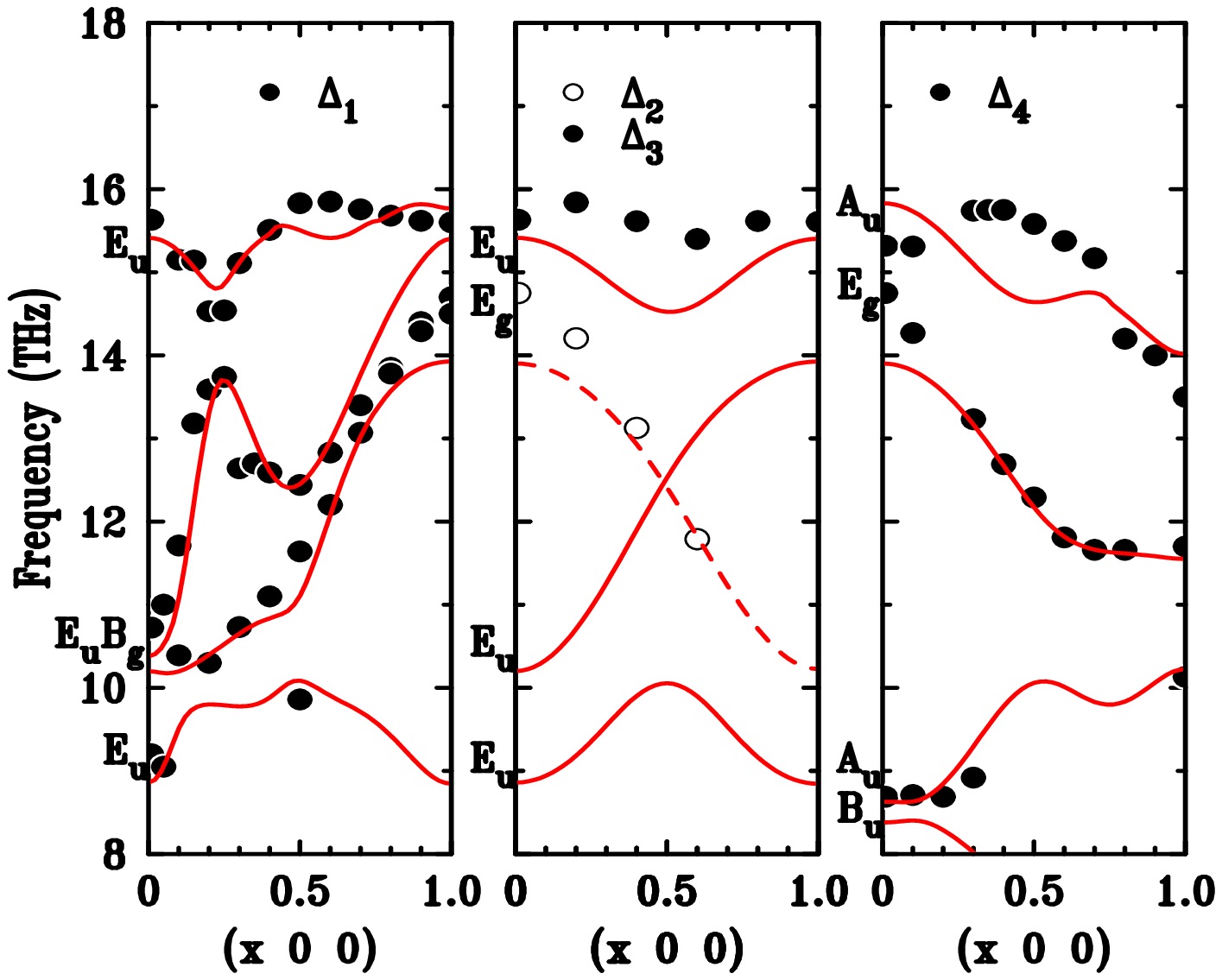}
} \caption{ Dispersion of the high-energy phonon branches in
\nccoef ~ in [100] direction. Symbols denote the experimental data
and lines frequencies calculated with the phenomenological model.
}\label{dispx00}
\end{figure}

\begin{figure}[tp]
\resizebox{0.6\figwidth}{!}{
\includegraphics*{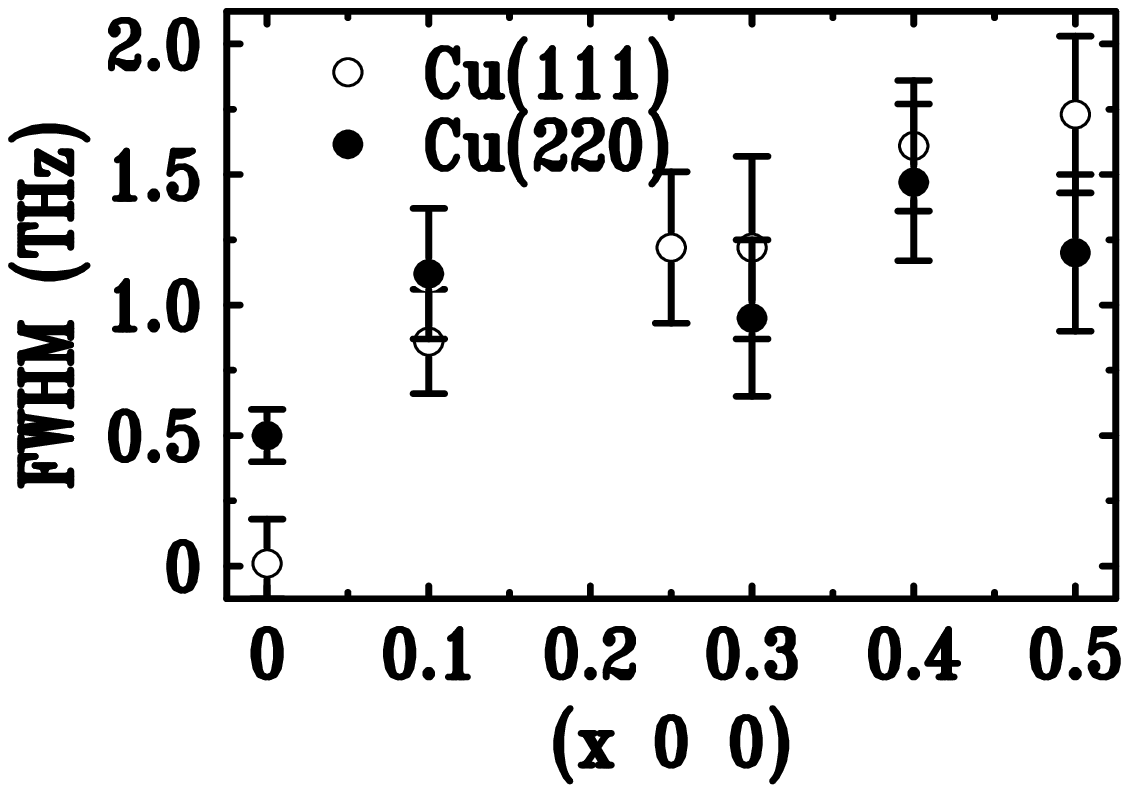}
} \caption{ Resolution corrected linewidths observed for
plane-polarized bond-stretching modes in \nccoef ~ in the [100]
direction at T = 10 K.}\label{widths}
\end{figure}

Energy scans at Q-values with favorable structure factors for
Cu-O1-bond-stretching phonons are shown in Fig. \ref{scanx00}.
Results are shown both for medium and high resolution set-ups. At
the zone center, (3 1 0), the bond-stretching mode can be easily
detected. For intermediate \vq -values, however, it is obscured
by the strong signal of the Nd-O2 vibrations, whose frequency
rapidly increases with $q$ whereas frequencies associated with the
Cu-O1-bond-stretching vibrations abruptly decrease. The steep
frequency increase in the O2-branch and the frequency decrease of
the bond-stretching modes render the measurement very sensitive
to the resolution conditions. At \vQ=(3.15 0 0) the increasing
branch is focused and at \vQ =(2.85 0 0 ) the decreasing one, see
Fig. \ref{scanx00}. The interaction of the O1 and the O2-branch
occurs around \vq =(0.2\ 0\ 0). At \vq =(0.3\ 0\ 0) the energy of
the Cu-O1-bond-stretching mode is already more than 2\ THz below
that of the O2-mode. In addition, the Cu-O1-bond-stretching mode
is broadened. Therefore, it has a stronger weight in the lower
resolution configuration. At \vq =(0.5 0 0) the bond-stretching
mode is clearly observed at 12.5\ THz. The dispersion of the \De
-modes is resumed in Fig. \ref{dispx00}.

We find a strong broadening of the Cu-O1-bond-stretching modes for
\vq -values in the Brillouin-zone. The full widths at half
maximum were obtained by fitting Gaussian profiles to the scans
and by correcting for the resolution, see Fig. \ref{widths}. Note
that the experimental resolution for measurements with the Cu(220)
monochromator is much smaller (0.6 THz) than the width of the
observed phonon peak. In spite of the difficulties to separate
Cu-O bond-stretching modes from other ones  our data
unambiguously shows that these modes are significantly broadened,
yielding another evidence for strong electron phonon coupling in
\ncco .

\begin{figure}[tp]
\resizebox{0.45\figwidth}{!}{
\includegraphics*{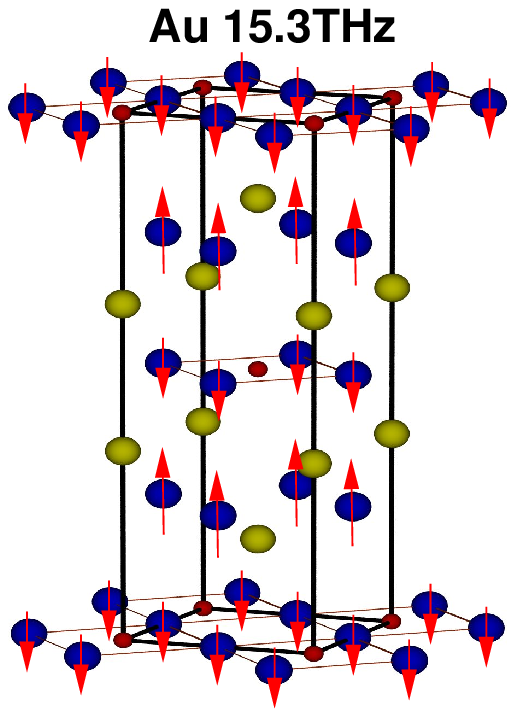}
} \resizebox{0.45\figwidth}{!}{
\includegraphics*{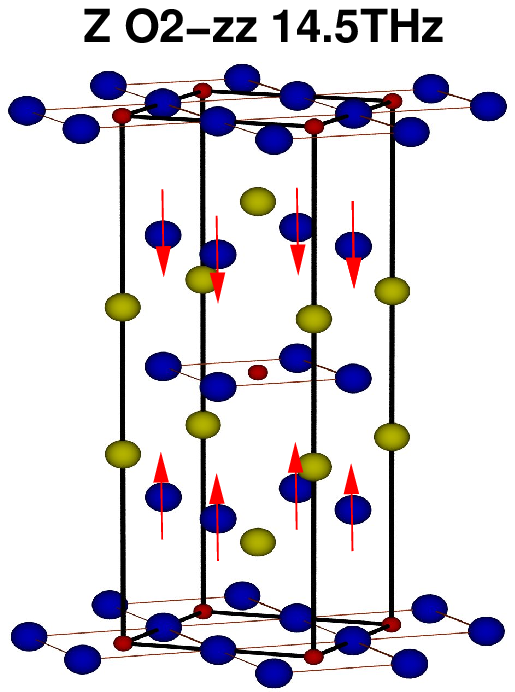}
} \resizebox{0.45\figwidth}{!}{
\includegraphics*{bg-10p3.eps}
} \resizebox{0.45\figwidth}{!}{
\includegraphics*{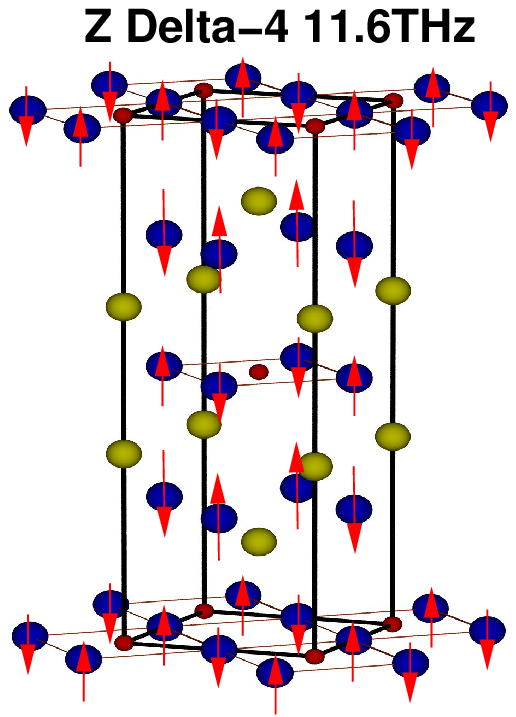}
} \caption{ Polarization patterns of zone-center and \vZ -point
modes with strong polarization along the $c$-direction. Large
blue balls correspond to oxygen atom, small red balls to Cu atoms
and the large yellow ones to Nd/Ce-sites. }\label{polpatt-z}
\end{figure}

Due to the body centered stacking of the CuO$_2$-planes in \ncco
, the reciprocal lattice vector \vZ =(1 0 0) is not a
Brillouin-zone centre but a zone boundary: the corresponding
modes are characterized by a phase shift of the displacements of
atoms at $\mathbf{r}$=(0 0 0) and at (0.5 0.5 0.5). Therefore,
the displacement patterns of the modes at the zone center and of
those at \vZ \ just differ by a phase shift in neighboring CuO$_2$
planes.  For the bond-stretching modes, one may expect similar
phonon frequencies at $\Gamma$ and at \vZ \ due to the weakness
of the inter-plane couplings involved. Following the Cu-O1
bond-stretching signal beyond (0.5 0 0) we indeed find the
expected frequency increase. However, between (0.6 0 0) and (1 0
0) there is an interaction with another branch. The \De -branch
starting at the $B_g$-mode exhibits an increasing dispersion.
Beyond \vq =(0.6 0 0) these modes strongly mix with the
Cu-O1-bond-stretching ones. The high-energy  phonon dispersion
obtained by measurements in different Brillouin-zones, including
zones with a large \vQ $_l$ component, are resumed in Fig.
\ref{dispx00}. Only a few measurements were made in the low
energy region because we did not expect strong doping-induced
changes here. The zone center modes in the range 8.5 to 10.5 THz
were investigated in some detail because there is a strong
interaction between $\Delta_1$ branches of different polarization
starting from these modes. These interactions virtually vanish,
however, for \vq $>$ 0.1 because the dispersion of the
interacting modes is very different.

\begin{figure}[tp]
\resizebox{0.8\figwidth}{!}{
\includegraphics*{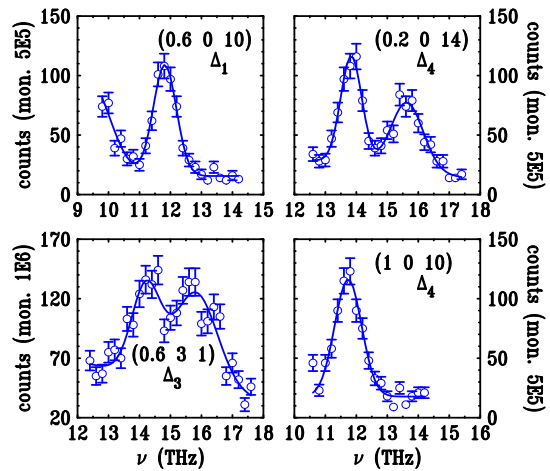}
} \caption{ Several scans across high-energy phonons with
different characters as indicated in the labels. Symbols denote
the raw intensities and  lines correspond to fits with Gaussians.
}\label{scanx00z}
\end{figure}

We now turn to the $c$-polarized high-energy modes. Their
polarization patterns  are shown in Fig. \ref{polpatt-z}:  at the
zone center, there is an $A_u$-mode near 15\ THz and the
$B_g$-mode at 10.3\ THz already mentioned. The O2-atoms exhibit a
polarization along the $c$ direction in both modes, but the modes
differ with respect to the phase between neighboring O2-sites. In
the $A_u$-mode, all O2's move in phase, whereas they move out of
phase in the $B_g$-mode, see Fig. \ref{polpatt-z}. At the \vZ
-point, too, there are two $c$-polarized O2-modes corresponding to
the in-phase and out-of-phase movement of the O2-ions at the same
$z$-level. The $B_g$ mode is connected through the \de -branches
with the in-phase movement at higher frequency; note that there
are some interactions with other branches. The $A_u$-mode
connects with the in-phase \vZ -mode through the $\Delta
_4$-branches. Typical scans aiming at the frequencies of these
phonon modes are shown in Fig. \ref{scanx00z}. Compared to the
longitudinal plane-polarized modes these phonons are better
defined, i.e. their line-widths are smaller. The in-phase \de
-mode at \vZ ~ is of special interest since all O2-sites move
in-phase, which means that all neighboring O2's of a particular
CuO$_2$-layer come closer at the same time, see Fig.
\ref{polpatt-z} . Therefore, this O2-vibration may induce strong
charge fluctuations between the CuO$_2$ planes; it is related in
character with the half-breathing mode, and one might expect a
strong coupling with an inter-layer charge modulation. The
corresponding mode in \lscofz , labeled $O_{ZZ}$, indeed is rather
anomalous with a large line width and a frequency renormalisation
compared to the parent compound by several \ THz \cite{5}. This
effect is interpreted by Falter et al. \cite{falter} as due to
strong coupling with the inter-layer plasmon mode. We find that
the $O_{ZZ}$-mode is not well defined in \nccofz \ as well
indicating a large intrinsic line width. Unfortunately, it is
difficult to isolate the contributions of the $c$-polarized
Z-point phonon from those with the in-plane polarization , but
the observation of a three-peak structure obtained at several \vZ
-points clearly proves that the $c$-polarized mode exhibits a
high energy. There is strong evidence that the $O_{ZZ}$-mode has
a large line width. Compared to \lscofz ~ the $O_{ZZ}$-mode in
\nccofz ~ appears to be less anomalous.

\begin{figure}[tp]
\resizebox{0.9\figwidth}{!}{
\includegraphics*{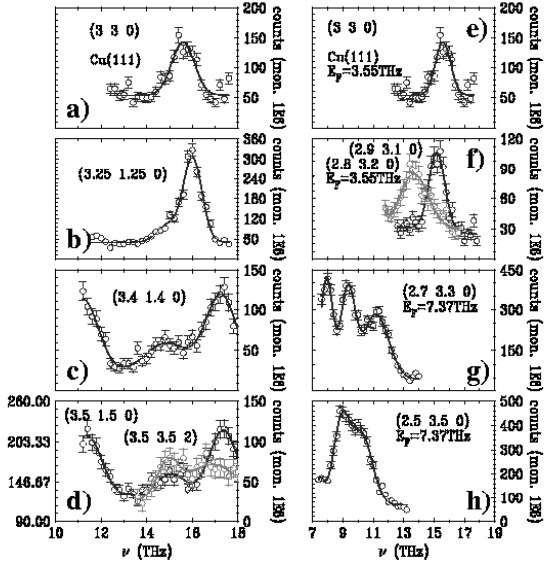}
}
\caption{ Scans across the longitudinal a-d) and transverse e-h)
bond-stretching modes with \se ~ and $\Sigma _3$ symmetry,
respectively. All scans were performed using a Cu(111)
monochromator with a final energy of $E_f$=3.55\ THz, except that
in g) and h) where $E_f$=7.37\ THz.
 .}\label{scanxx0}
\end{figure}

\subsection{III.2 Phonon dispersion in [110] direction }

\begin{figure}[tp]
\resizebox{0.95\figwidth}{!}{
\includegraphics*{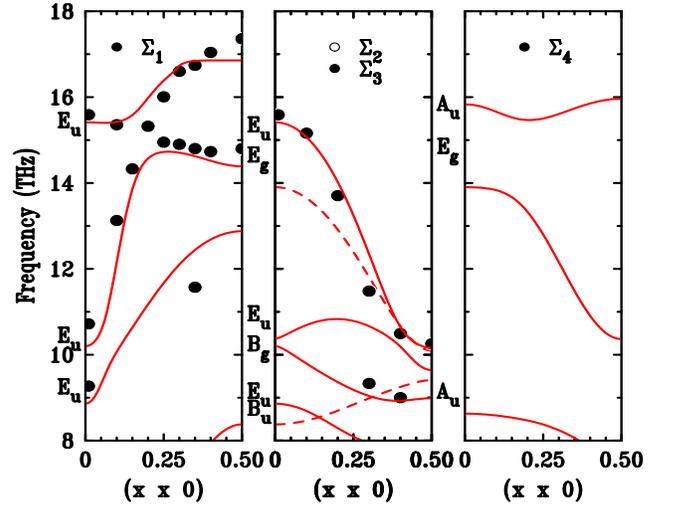}
} \caption{ Dispersion of the high-energy phonon branches in
\nccoef ~ in [110] direction. Symbols denote measured frequencies
and lines results of the lattics dynamical calculations (the
broken line in the middle corresponds to the $\Sigma _2$
representation.}\label{dispxx0}
\end{figure}

The phonon dispersion in the [110] direction, $\Sigma$, was
studied, again with the focus on the bond-stretching branches.
Typical scans are shown in Fig. \ref{scanxx0}, and the phonon
dispersion derived from the data is drawn in Fig. \ref{dispxx0}.
Again, the analysis is complicated by the interaction of the Cu-O1
with the Nd-O2 bond-stretching branches.  The scattering
associated with the latter modes is in most configurations
dominant. Nevertheless, we were able to unambigously determine the
dispersion of the branches of interest. The following picture
emerged: the branch associated with Nd-O2 vibrations disperses
steeply upwards from its starting point at $\nu$ = 9.4\ THz and
becomes the highest one in frequency for \vq \ larger than (0.15
0.15 0). At the zone boundary it reaches a frequency of 17.4\ THz,
the highest in the entire dispersion. Raman experiments have
observed a peak at just this frequency \cite{31} suggesting that
it should be attributed to the (0.5 0.5 0) LO O2-mode. This mode,
which is not Raman active, may become so through the disorder
induced by the Ce-doping. However, it remains astonishing that
this disorder induced Raman-peak is that strong. In contrast to
the Nd-O2 branch, the Cu-O1-bond-stretching branch is found to
exhibit a downward dispersion similar to the observation along the
[100] direction. However, the downward dispersion is less
pronounced along [110] and has a more conventional shape, i.e. it
is sine-like.

The transverse Cu-O1-bond-stretching branch of $\Sigma
_3$-symmetry also exhibits a steep frequency decrease when passing
into the Brillouin-zone, see the right part of Fig. \ref{scanxx0}
and Fig. \ref{dispxx0}. The transverse bond-stretching branch
along the [110] direction ends at the zone-boundary in the
so-called quadrupolar mode, whose polarization pattern is given in
Fig. 1. Its low frequency compared to that of the zone centre
mode is perfectly described by the lattice dynamical model
without any need for a special force constant. The shape of this
branch has to be considered as  normal. A similar shape is
observed in many other perovskites. The relatively low frequency
of the quadrupolar mode can be understood from the fact that this
vibration involves weaker force constants compared to the zone
center bond-stretching mode, as the O1-O1 distances vary only
little.

\subsection{IV. Discussion of bond-stretching phonon anomalies }

\begin{figure}[tp]
\resizebox{0.5\figwidth}{!}{
\includegraphics*{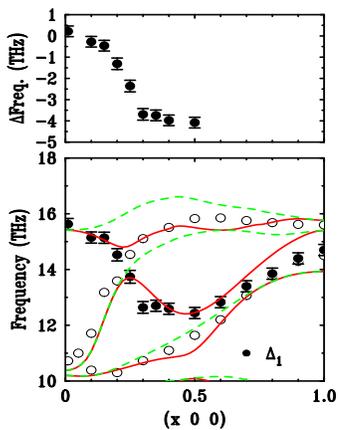}
} \caption{ (Below) Dispersion of the bond-stretching phonon
frequencies (filled circles) and other modes (open circles)
compared to the calculated phonon frequencies. The dashed and
continous lines correspond to screened shell-model calculations
without and with the additional breathing-force constants,
respectively. (Above) Difference between the measured
bond-stretching frequencies and the shell-model calculations
without the breathing forces. }\label{dispx00b}
\end{figure}

-- {\it Comparision with inelastic x-ray scattering data} --
Qualitatively, our results agree with the anticrossing scenario of
Cu-O1 bond-stretching and Nd-O2 modes which was deduced from an
inelastic x-ray experiment \cite{26,27}, but quantitatively, there
are significant differences for \vq -values away from the
Brilluoin-zone center.  For the [100] direction we find good
agreement up to \vq =(0.15 0 0), whereas already at \vq =(0.2 0 0)
the tentative identification of the bond-stretching mode near 12\
THz seems to be incorrect, see Fig. \ref{dispx00b} . The inelastic
x-ray spectra are dominated by the contribution of the Nd-O2 peak,
therefore, the bond-stretching frequencies could not be determined
unambiguously but had to be fitted to the little structured back
ground. For instance, inspection of Fig. 2 in \cite{26} shows that
no clear peak was observed at the position of the Cu-O1 mode for
\vq =(0.4 0 0). The approximate agreement between the x-ray and
the neutron frequencies for this \vq \ appears to be somewhat
accidental.

The problems encountered in the x-ray experiment aiming at the
dispersion of the Cu-O1 bond-stretching modes in the [110]
direction are similar to those just discussed for the [100]
direction. In particular, the inelastic structure factors of
these modes are very small for \vq \ beyond (0.25,0.25,0) in the
Brillouin zone chosen for the measurements. Therefore, it is not
surprising that the x-ray data for this q-region are quite
inaccurate and fall 1.5 THz below the neutron data.

In summary, our neutron results confirm the anti-crossing scenario
discussed in \cite{26,27} but only qualitatively so. The neutron
experiment is much better suited to resolve the complex
multi-branch dispersion by examining many different Brillouin
zones and due to the fact that oxygen and heavy-ion modes posses
similar dynamical structure factors for neutron scattering. Higher
order scattering events may play a stronger role in the case of
inelastic x-ray scattering studies on the oxygen modes.

-- {\it Comparision of bond-stretching phonon anomalies in high
temperature superconductors} -- Due to the mixing of several
branches in \nccoef ~ it is not obvious to isolate the behavior of
the bond-stretching modes in the [100] direction from Fig.
\ref{dispx00}. Therefore, we present the \De - data again in Fig.
\ref{dispx00b}, where the frequencies of the modes with
bond-stretching character are drawn as filled and all other
contributions as open circles, respectively. The separation can
be easily done in the first half of the Brillouin-zone, but is
less obvious in the second half due to the interaction with the
$c$-polarized modes, which is more extended in \vq -space. The
presentation in Fig. \ref{dispx00b} clearly shows the resemblance
of the bond-stretching dispersion in \nccoef ~ with that in
\lscoef \ \cite{6}. The observed dispersion is rather well
reproduced by the shell model after inclusion of the additional
breathing constants except that the frequency drop around \vq =
(0.25,0,0) is sharper in experiment . We emphasize that the
classical shell model without such force-constants predicts an
upwards bond-stretching dispersion as is indicated in Fig.
\ref{dispx00b} by broken lines. In the top of Fig. \ref{dispx00b}
we show the differences between the measured bond-stretching
frequencies and those calculated from the shell model without
breathing constants. These differences demonstrate the influence
of the anomalous strong  electron-phonon coupling in \ncco .

\begin{figure}[tp]
\resizebox{0.95\figwidth}{!}{
\includegraphics*[angle=90]{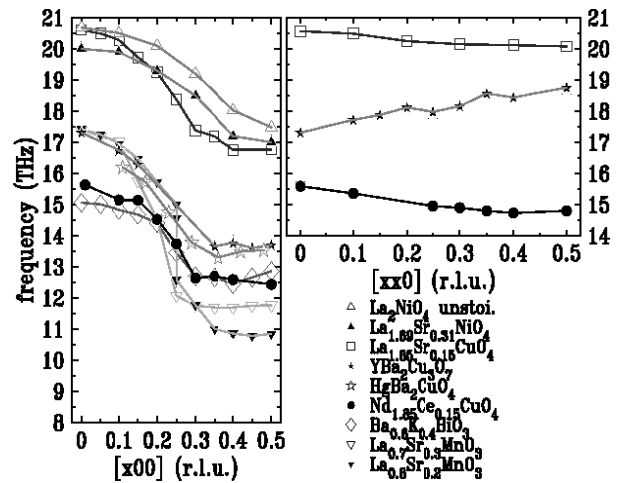}}
\caption{Comparison of the phonon anomaly in the bond-stretching
branches observed in different metallic oxide perovskite
materials. In the left we show the dispersion along the [100]
direction and in the right part that for the [110] direction.
Data were taken from references : Unstoichiometric \lanio \
\cite{15}, La$_{1.69}$Sr$_{0.31}$NiO$_4$ \ \cite{21}, \lscofz \
\cite{6}, \ybcos \ \cite{8}, \hgbco \ \cite{28}, \bkbo \
\cite{12}, and  \lsmo \ \cite{14}.}\label{dispall}
\end{figure}

The anomalous bond-stretching dispersion in \nccoef ~ is compared
with that of other cuprate superconductors in Fig. \ref{dispall}.
Evidently, the dispersion in the [100] direction is step-like in
all optimally doped cuprate superconductors investigated so far.
The different levels of bond-stretching phonon frequencies can be
attributed to different  Cu-O1-bond distances induced either
through different dopants or through internal strain. When
comparing the frequencies of the half-breathing mode to those of
the zone-centre bond-stretching mode we find the strongest
renormalisation in the two high-T$_c$ materials \ybco \ and
\hgbco  ; \lscoef ~ exhibits a slightly stronger softening than
\nccoef ~ along [100] \cite{kommentar2}.
In contrast, the downward dispersion along the [110] direction is
most pronounced in \nccoef  whereas the corresponding branch in
\lscoef ~ is nearly flat and  exhibits even a weak frequency
increase in \ybco \ \cite{5}.  That is to say,  \nccoef ~ shows
the least anisotropy between the frequency renormalisation along
the [100] and the [110] directions of all optimally doped
cuprates studied so far. In this sense, it can be said that the
phonon softening in \nccoef  \ has the least one-dimensional
character. It remains to be seen whether this fact is related to
the distinct character of the charge carriers, electron versus
hole-like, in the cuprate high-temperature superconductors. It is
worth mentioning that the one-dimensional nature of the
over-screening effect is most pronounced in the 90\
K-superconductor \ybcos . But we want to emphasize once more,
that the anomalous bond-stretching phonon dispersion in all
superconducting cuprates is astonishingly similar.

There is an on-going debate about the importance of electron
phonon coupling for the kink-like features observed in ARPES
spectra of many cuprate high temperature superconductors
\cite{shen,lanzarra,dordevic}. However, there seem to be
significant differences between the ARPES features of hole and
electron doped compounds \cite{armitage}, whereas we find the
phonon anomalies in the bond-stretching branches in all cuprates
to be essentially the same. This raises doubts about a dominant
role of the bond-stretching modes for the ARPES features.

--{\it Comparison with other metallic perovskites} -- The
over-screening effects in the bond-stretching dispersion are not
a particularity of the superconducting cuprates but are seen in
many doped perovskites : nickelates \cite{15,21}, manganates
\cite{14} and superconducting \bkbo \ \cite{12,19,20}, the only
metallic perovskite compound which does not exhibit the effect is
superconducting \srruo \ \cite{braden-unp}, allthough this
material is the best metal. The frequency softening is even much
more pronounced in manganates and bismuthates than in the cuprates
\cite{20}.

Fig. \ref{dispall} further demonstrates that many perovskite
materials exhibit even a similar shape of the bond-stretching
dispersion along [100]. The step-like shape of the dispersion is
observed in \bkbo ~ and in \lsmo \ for x=0.2 and 0.3 . In the
nickelates, however, the dispersion is nearly sinusoidal. This
dispersion may be related to the insulating electronic behavior
in nickelates \cite{20,imada}. The absolute size of the
over-screening effects differs essentially between the systems.
The effect along [100] is strongest in the manganates where it
amounts to a reduction by 30\% . Along the other directions in
reciprocal space, the manganates and  \bkbo ~ exhibit
over-screening effects which even exceed those occurring along
[100]. For instance, the volume breathing mode in \bkbo ~ is
renormalized by 40\% compared to the insulating reference
material \cite{19,20}. A strong softening in the [110] direction
was observed also in the nickelates. This indicates that the
electron-lattice coupling is quite strong in these  materials,
even considerably stronger than in the cuprates. This conclusion
is corroborated by the fact that in the manganites and in  \bkbo
~  the phonon anomalies are not restricted to a line in
reciprocal space but are very strong also in other directions.
Therefore, they entail a much stronger shift of spectral weight in
the PDOS.

We think that there is a close relationship of the
bond-stretching phonon anomalies with the charge ordering
phenomena in each particular system \cite{20}. All these systems
including the cuprates are close to charge ordering
instabilities, which may be stabilized by a particular doping
\cite{imada,23,24}. Roughly speaking, those bond-stretching modes
are most renormalized in frequency whose polarization patterns
corresponds to the structural distortion induced by the charge
ordering. The most obvious difference between the cuprates and
other perovskite compounds consists in the fact that  the [100]
direction plays a special role only in the cuprates, what
perfectly reflects the one-dimensional nature of charge ordering
in the form of stripes in hole-doped cuprates.  In contrast,
charge ordering phenomena in nickelates, manganates and
bismuthates occur with a checker-board arrangement. Concurrently,
the strongest bond-stretching phonon anomalies in these compounds
occur along the related [110] or [111]-directions. The relation
between charge ordering and bond-stretching phonon anomalies also
holds for the ruthenate, since in layered ruthenates, which
exhibit a normal bond-stretching phonon dispersion, in-plane
charge ordering seems not to occur.

Charge ordering phenomena have so far not been observed in the
electron-doped cuprates. From their similar phonon dispersion,
one may expect them to appear most likely again along the
[100]-direction.

Anomalous peak splitting has been reported for a number of
perovskite materials and have been taken as fingerprints of an
inhomogeneous charge distribution, i.e. in \bkbo \ \cite{19,20},
in the nickelates \cite{21}, and also in \ybcos \ and \lscoef \
\cite{7,9}. For the latter two materials, however, it was shown
that there are no indications for intrinsic splittings of
bond-stretching branches \cite{10,6}. We note that in \nccoef \ as
well, all observed peak splittings of bond-stretching phonons can
be fully explained with standard theory of lattice dynamics, i.e.
due to the interaction of branches having the same symmetry.

\section{ IV. Conclusion }

Inelastic neutron scattering allowed us to quantitatively
characterize the high-energy phonon dispersion in \nccoef ~ .
Previous inelastic x-ray measurements had already shown that the
frequencies of the Cu-O bond-stretching branches are strongly
renormalized when compared to those of the insulating parent
compounds but reliable data could be obtained only for a limited
range of wave vectors very close to the zone center whereas the
frequencies of the linear and of the planar breathing modes were
not accessible.

The dispersion of the branches with Cu-O1 bond-stretching
character exhibits anomalies astonishingly similar to the
observations in hole doped cuprate superconductors. Along the
[100] direction the frequency of the bond-stretching modes
abruptly decreases near \vq =(0.25 0 0). This effect is slightly
weaker than in the hole doped materials. Moreover, these phonons
acquire a large linewidth which underpins their anomalous
character.  The longitudinal bond-stretching modes exhibit a
downward dispersion also along the [110] direction with a
frequency drop between zone center and zone boundary of about 1
THz but the dispersion shape of the branch appears to be more
normal. The softening of the planar breathing-mode softening in
\nccoef ~ is stronger than the corresponding effect in the hole
doped materials, in particular compared to \ybcos , indicating
that the anomalous softening of the linear breathing mode has
weaker one-dimensional character. However, the differences in the
phonon anomalies between hole and electron doped cuprates concern
details; the signatures of electron phonon coupling in the
bond-stretching phonons are found to be essentially the same in
all cuprate high temperature superconductors studied so far.

%
%

{\bf Acknowledgments}  This work was supported by the Deutsche
Forschungsgemeinschaft through the Sonderforschungsbereich 608
and by the specific grants-in-aid from the MEXT (Japan). We
acknowledge helpful discussions with M. d'Astuto.


\begin{references}
\bibitem{kommentar1} We have chosen a large value of the linear breathing constant
in order to describe the phonon dispersion but this value yields a
shift of some breathing-mode character towards the next lowest
branch. Nearly the complete bond-stretching character in \nccofz \
at \vq =(0.5,0,0) , however, appears to reside in the mode at
12.5THz.
\bibitem{kommentar2} In the case of \ncco ~ it is difficult to
directly compare undoped and doped compounds, because Cu-O1
bond-stretching modes are not pure in character in the undoped
material but exhibit significant Nd-O2 character.
\end{references}
\end{document}